\documentclass[11pt]{article}

\ifx\undefined\psfig\else \fi

\edef\psfigRestoreAt{\catcode`@=\number\catcode`@\relax}
\catcode`\@=11\relax
\newwrite\@unused
\def\typeout#1{{\let\protect\string\immediate\write\@unused{#1}}}
\def\figurepath{./}

\def\@nnil{\@nil}
\def\@empty{}
\def\@psdonoop#1\@@#2#3{}
\def\@psdo#1:=#2\do#3{\edef\@psdotmp{#2}\ifx\@psdotmp\@empty \else
    \expandafter\@psdoloop#2,\@nil,\@nil\@@#1{#3}\fi}
\def\@psdoloop#1,#2,#3\@@#4#5{\def#4{#1}\ifx #4\@nnil \else
       #5\def#4{#2}\ifx #4\@nnil \else#5\@ipsdoloop #3\@@#4{#5}\fi\fi}
\def\@ipsdoloop#1,#2\@@#3#4{\def#3{#1}\ifx #3\@nnil 
       \let\@nextwhile=\@psdonoop \else
      #4\relax\let\@nextwhile=\@ipsdoloop\fi\@nextwhile#2\@@#3{#4}}
\def\@tpsdo#1:=#2\do#3{\xdef\@psdotmp{#2}\ifx\@psdotmp\@empty \else
    \@tpsdoloop#2\@nil\@nil\@@#1{#3}\fi}
\def\@tpsdoloop#1#2\@@#3#4{\def#3{#1}\ifx #3\@nnil 
       \let\@nextwhile=\@psdonoop \else
      #4\relax\let\@nextwhile=\@tpsdoloop\fi\@nextwhile#2\@@#3{#4}}
\newread\ps@stream
\newif\ifnot@eof       
\newif\if@noisy        
\newif\if@atend        
\newif\if@psfile       
{\catcode`\%=12\global\gdef\epsf@start{
\def\epsf@PS{PS}
\def\epsf@getbb#1{%
\openin\ps@stream=#1
\ifeof\ps@stream\typeout{Error, File #1 not found}\else
   {\not@eoftrue \chardef\other=12
    \def\do##1{\catcode`##1=\other}\dospecials \catcode`\ =10
    \loop
       \if@psfile
	  \read\ps@stream to \epsf@fileline
       \else{
	  \obeyspaces
          \read\ps@stream to \epsf@tmp\global\let\epsf@fileline\epsf@tmp}
       \fi
       \ifeof\ps@stream\not@eoffalse\else
       \if@psfile\else
       \expandafter\epsf@test\epsf@fileline:. \\%
       \fi
          \expandafter\epsf@aux\epsf@fileline:. \\%
       \fi
   \ifnot@eof\repeat
   }\closein\ps@stream\fi}%
\long\def\epsf@test#1#2#3:#4\\{\def\epsf@testit{#1#2}
			\ifx\epsf@testit\epsf@start\else
\typeout{Warning! File does not start with `\epsf@start'.  It may not be a PostScript file.}
			\fi
			\@psfiletrue} 
{\catcode`\%=12\global\let\epsf@percent=
\long\def\epsf@aux#1#2:#3\\{\ifx#1\epsf@percent
   \def\epsf@testit{#2}\ifx\epsf@testit\epsf@bblit
	\@atendfalse
        \epsf@atend #3 . \\%
	\if@atend	
	   \if@verbose{
		\typeout{psfig: found `(atend)'; continuing search}
	   }\fi
        \else
        \epsf@grab #3 . . . \\%
        \not@eoffalse
        \global\no@bbfalse
        \fi
   \fi\fi}%
\def\epsf@grab #1 #2 #3 #4 #5\\{%
   \global\def\epsf@llx{#1}\ifx\epsf@llx\empty
      \epsf@grab #2 #3 #4 #5 .\\\else
   \global\def\epsf@lly{#2}%
   \global\def\epsf@urx{#3}\global\def\epsf@ury{#4}\fi}%
\def\epsf@atendlit{(atend)} 
\def\epsf@atend #1 #2 #3\\{%
   \def\epsf@tmp{#1}\ifx\epsf@tmp\empty
      \epsf@atend #2 #3 .\\\else
   \ifx\epsf@tmp\epsf@atendlit\@atendtrue\fi\fi}

\chardef\letter = 11
\chardef\other = 12

\newif \ifdebug 
\newif\ifc@mpute 
\c@mputetrue 

\let\then = \relax
\def\r@dian{pt }
\let\r@dians = \r@dian
\let\dimensionless@nit = \r@dian
\let\dimensionless@nits = \dimensionless@nit
\def\internal@nit{sp }
\let\internal@nits = \internal@nit
\newif\ifstillc@nverging
\def \Mess@ge #1{\ifdebug \then \message {#1} \fi}

{ 
	\catcode `\@ = \letter
	\gdef \nodimen {\expandafter \n@dimen \the \dimen}
	\gdef \term #1 #2 #3%
	       {\edef \t@ {\the #1}
		\edef \t@@ {\expandafter \n@dimen \the #2\r@dian}%
		\t@rm {\t@} {\t@@} {#3}%
	       }
	\gdef \t@rm #1 #2 #3%
	       {{%
		\count 0 = 0
		\dimen 0 = 1 \dimensionless@nit
		\dimen 2 = #2\relax
		\Mess@ge {Calculating term #1 of \nodimen 2}%
		\loop
		\ifnum	\count 0 < #1
		\then	\advance \count 0 by 1
			\Mess@ge {Iteration \the \count 0 \space}%
			\Multiply \dimen 0 by {\dimen 2}%
			\Mess@ge {After multiplication, term = \nodimen 0}%
			\Divide \dimen 0 by {\count 0}%
			\Mess@ge {After division, term = \nodimen 0}%
		\repeat
		\Mess@ge {Final value for term #1 of 
				\nodimen 2 \space is \nodimen 0}%
		\xdef \Term {#3 = \nodimen 0 \r@dians}%
		\aftergroup \Term
	       }}
	\catcode `\p = \other
	\catcode `\t = \other
	\gdef \n@dimen #1pt{#1} 
}

\def \Divide #1by #2{\divide #1 by #2} 

\def \Multiply #1by #2
       {{
	\count 0 = #1\relax
	\count 2 = #2\relax
	\count 4 = 65536
	\Mess@ge {Before scaling, count 0 = \the \count 0 \space and
			count 2 = \the \count 2}%
	\ifnum	\count 0 > 32767 
	\then	\divide \count 0 by 4
		\divide \count 4 by 4
	\else	\ifnum	\count 0 < -32767
		\then	\divide \count 0 by 4
			\divide \count 4 by 4
		\else
		\fi
	\fi
	\ifnum	\count 2 > 32767 
	\then	\divide \count 2 by 4
		\divide \count 4 by 4
	\else	\ifnum	\count 2 < -32767
		\then	\divide \count 2 by 4
			\divide \count 4 by 4
		\else
		\fi
	\fi
	\multiply \count 0 by \count 2
	\divide \count 0 by \count 4
	\xdef \product {#1 = \the \count 0 \internal@nits}%
	\aftergroup \product
       }}

\def\r@duce{\ifdim\dimen0 > 90\r@dian \then   
		\multiply\dimen0 by -1
		\advance\dimen0 by 180\r@dian
		\r@duce
	    \else \ifdim\dimen0 < -90\r@dian \then  
		\advance\dimen0 by 360\r@dian
		\r@duce
		\fi
	    \fi}

\def\Sine#1%
       {{%
	\dimen 0 = #1 \r@dian
	\r@duce
	\ifdim\dimen0 = -90\r@dian \then
	   \dimen4 = -1\r@dian
	   \c@mputefalse
	\fi
	\ifdim\dimen0 = 90\r@dian \then
	   \dimen4 = 1\r@dian
	   \c@mputefalse
	\fi
	\ifdim\dimen0 = 0\r@dian \then
	   \dimen4 = 0\r@dian
	   \c@mputefalse
	\fi
	\ifc@mpute \then
		\divide\dimen0 by 180
		\dimen0=3.141592654\dimen0
		\dimen 2 = 3.1415926535897963\r@dian 
		\divide\dimen 2 by 2 
		\Mess@ge {Sin: calculating Sin of \nodimen 0}%
		\count 0 = 1 
		\dimen 2 = 1 \r@dian 
		\dimen 4 = 0 \r@dian 
		\loop
			\ifnum	\dimen 2 = 0 
			\then	\stillc@nvergingfalse 
			\else	\stillc@nvergingtrue
			\fi
			\ifstillc@nverging 
			\then	\term {\count 0} {\dimen 0} {\dimen 2}%
				\advance \count 0 by 2
				\count 2 = \count 0
				\divide \count 2 by 2
				\ifodd	\count 2 
				\then	\advance \dimen 4 by \dimen 2
				\else	\advance \dimen 4 by -\dimen 2
				\fi
		\repeat
	\fi		
			\xdef \sine {\nodimen 4}%
       }}

\def\Cosine#1{\ifx\sine\UnDefined\edef\Savesine{\relax}\else
		             \edef\Savesine{\sine}\fi
	{\dimen0=#1\r@dian\multiply\dimen0 by -1
	 \advance\dimen0 by 90\r@dian
	 \Sine{\nodimen 0}
	 \xdef\cosine{\sine}
	 \xdef\sine{\Savesine}}}	      

\def\psdraft{
	\def\@psdraft{0}
}
\def\psfull{
	\def\@psdraft{100}
}

\psfull

\newif\if@draftbox
\def\psnodraftbox{
	\@draftboxfalse
}
\@draftboxtrue

\newif\if@prologfile
\newif\if@postlogfile
\def\pssilent{
	\@noisyfalse
}
\def\psnoisy{
	\@noisytrue
}
\psnoisy
\newif\if@bbllx
\newif\if@bblly
\newif\if@bburx
\newif\if@bbury
\newif\if@height
\newif\if@width
\newif\if@rheight
\newif\if@rwidth
\newif\if@angle
\newif\if@clip
\newif\if@verbose
\newif\if@scale
\def\@p@@sclip#1{\@cliptrue}

\def\@p@@sfile#1{\def\@p@sfile{null}%
	        \openin1=#1
		\ifeof1\closein1%
		       \openin1=\figurepath#1
			\ifeof1\typeout{Error, File #1 not found}
			   \if@bbllx\if@bblly\if@bburx\if@bbury
			      \def\@p@sfile{#1}%
			   \fi\fi\fi\fi
			\else\closein1
			    \edef\@p@sfile{\figurepath#1}%
                        \fi%
		 \else\closein1%
		       \def\@p@sfile{#1}%
		 \fi}
\def\@p@@sfigure#1{\def\@p@sfile{null}%
	        \openin1=#1
		\ifeof1\closein1%
		       \openin1=\figurepath#1
			\ifeof1\typeout{Error, File #1 not found}
			   \if@bbllx\if@bblly\if@bburx\if@bbury
			      \def\@p@sfile{#1}%
			   \fi\fi\fi\fi
			\else\closein1
			    \def\@p@sfile{\figurepath#1}%
                        \fi%
		 \else\closein1%
		       \def\@p@sfile{#1}%
		 \fi}

\def\@p@@sbbllx#1{
		\@bbllxtrue
		\dimen100=#1
		\edef\@p@sbbllx{\number\dimen100}
}
\def\@p@@sbblly#1{
		\@bbllytrue
		\dimen100=#1
		\edef\@p@sbblly{\number\dimen100}
}
\def\@p@@sbburx#1{
		\@bburxtrue
		\dimen100=#1
		\edef\@p@sbburx{\number\dimen100}
}
\def\@p@@sbbury#1{
		\@bburytrue
		\dimen100=#1
		\edef\@p@sbbury{\number\dimen100}
}
\def\@p@@sheight#1{
		\@heighttrue
		\dimen100=#1
   		\edef\@p@sheight{\number\dimen100}
}
\def\@p@@swidth#1{
		\@widthtrue
		\dimen100=#1
		\edef\@p@swidth{\number\dimen100}
}
\def\@p@@srheight#1{
		\@rheighttrue
		\dimen100=#1
		\edef\@p@srheight{\number\dimen100}
}
\def\@p@@srwidth#1{
		\@rwidthtrue
		\dimen100=#1
		\edef\@p@srwidth{\number\dimen100}
}
\def\@p@@sangle#1{
		\@angletrue
		\edef\@p@sangle{#1} 
}
\def\@p@@ssilent#1{ 
		\@verbosefalse
}
\def\@p@@sscale#1{
		\def\@p@scale{#1}
		\@scaletrue
}
\def\@p@@sprolog#1{\@prologfiletrue\def\@prologfileval{#1}}
\def\@p@@spostlog#1{\@postlogfiletrue\def\@postlogfileval{#1}}
\def\@cs@name#1{\csname #1\endcsname}
\def\@setparms#1=#2,{\@cs@name{@p@@s#1}{#2}}
\def\ps@init@parms{
		\@bbllxfalse \@bbllyfalse
		\@bburxfalse \@bburyfalse
		\@heightfalse \@widthfalse
		\@rheightfalse \@rwidthfalse
		\@scalefalse
		\def\@p@sbbllx{}\def\@p@sbblly{}
		\def\@p@sbburx{}\def\@p@sbbury{}
		\def\@p@sheight{}\def\@p@swidth{}
		\def\@p@srheight{}\def\@p@srwidth{}
		\def\@p@sangle{0}
		\def\@p@sfile{}
		\def\@p@scost{10}
		\def\@sc{}
		\@prologfilefalse
		\@postlogfilefalse
		\@clipfalse
		\if@noisy
			\@verbosetrue
		\else
			\@verbosefalse
		\fi
}
\def\parse@ps@parms#1{
	 	\@psdo\@psfiga:=#1\do
		   {\expandafter\@setparms\@psfiga,}}
\newif\ifno@bb
\def\bb@missing{
	\if@verbose{
		\typeout{psfig: searching \@p@sfile \space  for bounding box}
	}\fi
	\no@bbtrue
	\epsf@getbb{\@p@sfile}
        \ifno@bb \else \bb@cull\epsf@llx\epsf@lly\epsf@urx\epsf@ury\fi
}	
\def\bb@cull#1#2#3#4{
	\dimen100=#1 bp\edef\@p@sbbllx{\number\dimen100}
	\dimen100=#2 bp\edef\@p@sbblly{\number\dimen100}
	\dimen100=#3 bp\edef\@p@sbburx{\number\dimen100}
	\dimen100=#4 bp\edef\@p@sbbury{\number\dimen100}
	\no@bbfalse
}

\newdimen\p@intvaluex
\newdimen\p@intvaluey
\newdimen\@ffsetvalue
\newdimen\x@ffsetvalue
\newdimen\y@ffsetvalue

\def\compute@offset#1#2{{\dimen0=#1 sp\dimen1=#2 sp
			\advance\dimen1 by -\dimen0
			\dimen1=\sine\dimen1
			\dimen0=\cosine\dimen1
			\ifdim\dimen0<0sp \dimen1=0sp \fi
			\global\@ffsetvalue=\dimen1}}

\def\rotate@#1#2{{\dimen0=#1 sp\dimen1=#2 sp
		  \global\p@intvaluex=\cosine\dimen0
		  \dimen3=\sine\dimen1
		  \global\advance\p@intvaluex by -\dimen3
		  \global\p@intvaluey=\sine\dimen0
		  \dimen3=\cosine\dimen1
		  \global\advance\p@intvaluey by \dimen3
		  }}
\def\compute@bb{
		\no@bbfalse
		\if@bbllx \else \no@bbtrue \fi
		\if@bblly \else \no@bbtrue \fi
		\if@bburx \else \no@bbtrue \fi
		\if@bbury \else \no@bbtrue \fi
		\ifno@bb \bb@missing \fi
		\ifno@bb \typeout{FATAL ERROR: no bb supplied or found}
			\no-bb-error
		\fi
		\if@angle 
			\Sine{\@p@sangle}\Cosine{\@p@sangle}
			\compute@offset{\@p@sbblly}{\@p@sbbury}
			\x@ffsetvalue=\@ffsetvalue
			\compute@offset{\@p@sbburx}{\@p@sbbllx}
			\y@ffsetvalue=\@ffsetvalue

			\rotate@{\@p@sbbllx}{\@p@sbblly}
			\advance\p@intvaluex by -\x@ffsetvalue
			\advance\p@intvaluey by -\y@ffsetvalue
			\edef\@p@sbbllx{\number\p@intvaluex}
			\edef\@p@sbblly{\number\p@intvaluey}

			\rotate@{\@p@sbburx}{\@p@sbbury}
			\advance\p@intvaluex by \x@ffsetvalue
			\advance\p@intvaluey by \y@ffsetvalue
			\edef\@p@sbburx{\number\p@intvaluex}
			\edef\@p@sbbury{\number\p@intvaluey}
			{
			 \count0=\@p@sbbllx \count1=\@p@sbblly
		 	 \count2=\@p@sbburx \count3=\@p@sbbury
			 \dimen0=\@p@sbbllx sp\dimen1=\@p@sbblly sp
		 	 \dimen2=\@p@sbburx sp\dimen3=\@p@sbbury sp
			 \dimen203=\dimen2 \advance\dimen203 by -\dimen0
			 \dimen204=\dimen3 \advance\dimen204 by -\dimen1
			 \ifdim\dimen203<0sp 
			      \count203=\count2 \count2=\count0 
			      \count0=\count203 
			      \global\edef\@p@sbbllx{\number\count0}
			      \global\edef\@p@sbburx{\number\count2}
			 \fi
			 \ifdim\dimen204<0sp 
			       \count204=\count3
			       \count3=\count1
			       \count1=\count204
			       \global\edef\@p@sbblly{\number\count1}
			       \global\edef\@p@sbbury{\number\count3}
			 \fi
			}
		\fi
		\count203=\@p@sbburx
		\count204=\@p@sbbury
		\advance\count203 by -\@p@sbbllx
		\advance\count204 by -\@p@sbblly
		\edef\@bbw{\number\count203}
		\edef\@bbh{\number\count204}
}
\def\in@hundreds#1#2#3{\count240=#2 \count241=#3
		     \count100=\count240	
		     \divide\count100 by \count241
		     \count101=\count100
		     \multiply\count101 by \count241
		     \advance\count240 by -\count101
		     \multiply\count240 by 10
		     \count101=\count240	
		     \divide\count101 by \count241
		     \count102=\count101
		     \multiply\count102 by \count241
		     \advance\count240 by -\count102
		     \multiply\count240 by 10
		     \count102=\count240	
		     \divide\count102 by \count241
		     \count200=#1\count205=0
		     \count201=\count200
			\multiply\count201 by \count100
		 	\advance\count205 by \count201
		     \count201=\count200
			\divide\count201 by 10
			\multiply\count201 by \count101
			\advance\count205 by \count201
		     \count201=\count200
			\divide\count201 by 100
			\multiply\count201 by \count102
			\advance\count205 by \count201
		     \edef\@result{\number\count205}
}
\def\@ScaleInHundreds#1{
		\in@hundreds{#1}{\@p@scale}{100}
		\edef#1{\@result}
}
\def\compute@wfromh{
		\in@hundreds{\@p@sheight}{\@bbw}{\@bbh}
		\edef\@p@swidth{\@result}
}
\def\compute@hfromw{
		\in@hundreds{\@p@swidth}{\@bbh}{\@bbw}
		\edef\@p@sheight{\@result}
}
\def\compute@handw{
		\if@height 
			\if@width
			\else
				\compute@wfromh
			\fi
		\else 
			\if@width
				\compute@hfromw
			\else
				\edef\@p@sheight{\@bbh}
				\edef\@p@swidth{\@bbw}
			\fi
		\fi
}
\def\compute@resv{
		\if@rheight \else \edef\@p@srheight{\@p@sheight} \fi
		\if@rwidth \else \edef\@p@srwidth{\@p@swidth} \fi
}
\def\compute@sizes{
	\compute@bb
	\compute@handw
	\compute@resv
}
\def\psfig#1{\vbox {
	\ps@init@parms
	\parse@ps@parms{#1}
	\compute@sizes
	\if@scale
                \if@verbose
                        \typeout{psfig: scaling by \@p@scale}
                \fi
                \@ScaleInHundreds{\@p@swidth}
                \@ScaleInHundreds{\@p@sheight}
                \@ScaleInHundreds{\@p@srwidth}
                \@ScaleInHundreds{\@p@srheight}
        \fi
	\ifnum\@p@scost<\@psdraft{
		\if@verbose{
			\typeout{psfig: including \@p@sfile \space }
		}\fi
		\special{ps::[begin] 	\@p@swidth \space \@p@sheight \space
				\@p@sbbllx \space \@p@sbblly \space
				\@p@sbburx \space \@p@sbbury \space
				startTexFig \space }
		\if@angle
			\special {ps:: \@p@sangle \space rotate \space} 
		\fi
		\if@clip{
			\if@verbose{
				\typeout{(clip)}
			}\fi
			\special{ps:: doclip \space }
		}\fi
		\if@prologfile
		    \special{ps: plotfile \@prologfileval \space } \fi
		\special{ps: plotfile \@p@sfile \space }
		\if@postlogfile
		    \special{ps: plotfile \@postlogfileval \space } \fi
		\special{ps::[end] endTexFig \space }
		\vbox to \@p@srheight true sp{
			\hbox to \@p@srwidth true sp{
				\hss
			}
		\vss
		}
	}\else{
		\if@draftbox{		
			\hbox{\fbox{\vbox to \@p@srheight true sp{
			\vss
			\hbox to \@p@srwidth true sp{ \hss \@p@sfile \hss }
			\vss
			}}}
		}\else{
			\vbox to \@p@srheight true sp{
			\vss
			\hbox to \@p@srwidth true sp{\hss}
			\vss
			}
		}\fi

	}\fi
}}
\def\psglobal{\typeout{psfig: PSGLOBAL is OBSOLETE; use psprint -m instead}}
\psfigRestoreAt

\newif\ifpdf
\ifx\pdfoutput\undefined
  \pdffalse
\else
  \pdfoutput=1
  \pdftrue
\fi

\ifpdf
  \usepackage[pdftex]{graphicx}
  \usepackage[pdftex]{color}
  \DeclareGraphicsExtensions{.pdf,.png,.jpg}
\else
  \usepackage[dvips]{graphicx}
  \usepackage[dvips]{color}
  \DeclareGraphicsExtensions{.eps,.epsi,.ps}
\fi

\usepackage{times}

\def\midv{\mathop{\,|\,}}

\long\def\cbk#1{{\color{red}[CBK: #1]}}
\newlength\colwidth \setlength\colwidth{3.25in}

\title{Turning CARTwheels: \\An Alternating Algorithm 
for Mining Redescriptions}
\author{Deept Kumar$^\dagger$, Naren Ramakrishnan$^{\dagger *}$, Malcolm Potts$^\ddagger$, and Richard F. Helm$^\ddagger$\\
$^\dagger$Department of Computer Science\\
$^\ddagger$Virginia Tech Center for Genomics, Fralin Biotechnology Center\\
Virginia Tech, Blacksburg, VA 24061, USA\\
$^*$Contact: naren@cs.vt.edu}
\date{}
\begin{document}

\maketitle

\begin{abstract}
\noindent
We present an unusual algorithm involving classification trees 
where two trees are
grown in opposite directions so that they are matched at their leaves. This approach finds 
application in a new data mining task we formulate, called {\it redescription mining}. 
A redescription is a shift-of-vocabulary, or a different way of communicating information
about a given subset of data; the goal of redescription mining is to find subsets of 
data that afford multiple descriptions. We highlight the importance of this problem 
in domains such as bioinformatics, which exhibit an underlying richness and diversity
of data descriptors (e.g., genes can be studied in a variety of ways). Our approach
helps integrate multiple forms of characterizing datasets, situates the knowledge
gained from one dataset in the context of others, and harnesses
high-level abstractions for uncovering cryptic and subtle features of data.
Algorithm design decisions, implementation details, and experimental 
results are presented.
\end{abstract}
\thispagestyle{empty}

\section{Introduction}
Classification and regression trees (CART) were among the earliest proposed approaches
for pattern classification and data mining~\cite{cartbook}. While being powerful in 
terms of accuracy and efficiency
of induction, their results are also simple to understand as they mimic the 
decision-making logic of human experts. 
The renewed emphasis on data mining propagated by the knowledge discovery in 
databases (KDD) community in the early 1990s
has fueled a resurgence of interest in tree-based methods.
Researchers have revisited tree induction algorithms in the context of datasets
residing in secondary storage~\cite{ieee-scaling,dt-scaling}, creating scalable and 
highly efficient implementations~\cite{bradley-cacm}.
The many fielded applications of tree-based methods range from everyday uses 
such as spam filtering~\cite{hastie-book} to astrophysical domains such as classifying 
galaxies~\cite{kamath-galaxies}.

%
In this paper we introduce a new data mining task---{\it redescription mining}---and 
also propose a novel tree-based algorithm (CARTwheels) for mining redescriptions.  
A redescription is a shift-of-vocabulary, or a different way of communicating information
about a given subset of data; the goal of redescription mining is to find subsets of 
data that afford multiple descriptions. 

Consider the set of all countries in the world. The elements of this 
set can be described in various ways, e.g., geographical location, political status,
scientific capabilities, and economic prosperity.
Such descriptors
allow us to define various subsets of the given (universal) set. A redescription involves
a subset definable in two ways, for instance:
\vspace{-0.1in}

\begin{tabular}{lcl}\\ 
`Countries with $> 200$ Nobel prize winners' & $\Leftrightarrow$ & `Countries with
$> 150$ billionaires'\\
\end{tabular}

\vspace{0.1in}
\noindent
This redescription involves two descriptors, and says that the countries
with more than 200 Nobel prize winners are also those countries with more than
150 billionaires. One country satisfies both descriptors, namely U.S.A., and we
say that it has been redescribed. 
The strength of the redescription is given by the symmetric Jaccard's coefficient, which 
is the ratio
of the size of the intersection of two descriptors to the size of their union
(in this case, 1/1 = 1).

Descriptors on either side of a redescription can involve more than one entity, e.g.,

\vspace{-0.1in}
\begin{tabular}{lcl}\\
`Countries with defense budget $>$ \$30 billion' & $\Leftrightarrow$ & `Permanent members of 
U.N. Security Council'\\
\end{tabular}

\vspace{0.1in}
\noindent
This redescription is only approximate, however, 
since the left descriptor contains \{U.S.A., U.K., Japan, France, Germany\}
and the right descriptor represents \{U.S.A., U.K., Russia, France, China\}.
The Jaccard's coefficient is hence 3/7 = 0.428.

To strengthen redescriptions such as above, we can use more selective descriptors:

\vspace{-0.1in}
\begin{tabular}{lcl}\\
`Countries with declared nuclear arsenals' & $\Leftrightarrow$ & `Permanent members of 
U.N. Security Council'\\
\end{tabular}

\vspace{0.1in}
\noindent
which improves the Jaccard's coefficient to 5/8 = 0.625 since the left descriptor now
represents \{U.S.A., U.K., Russia, France, China, India, Israel, Pakistan\}.
Another approach to strengthening is to
form set-theoretic operations (union, intersection, difference) involving the 
given descriptors; e.g. the redescription

\vspace{-0.1in}
\begin{tabular}{lcl}\\
`Countries with defense budget $>$ \$30 billion' $\cap$ `Countries with declared nuclear arsenals' & $\Leftrightarrow$ & \\
\,\,\,\,\,\,\,\,\,\,\,\,`Permanent members of U.N. Security Council' $-$ `Countries with history of communism'& & \\
\end{tabular}

\vspace{0.1in}
\noindent
holds with 
Jaccard's coefficient $1$. It refers to three countries: \{U.S.A., U.K., France\}.

The inputs to redescription mining are the universal set of objects $O$ and 
two sets ($X$ and $Y$) of subsets of $O$. The elements of $X$ are the descriptors
$X_i$, and are assumed to form a covering of $O$ ($\bigcup_i X_i = O$). Similarly
$\bigcup_i Y_i = O$. The only requirement of a descriptor is that it be a proper
subset of $O$ and denote some logical grouping of the underlying objects (for ease
of interpretation). The goal of redescription mining is to find equivalence
relationships of the form $E \Leftrightarrow F$ that hold at or above a given
Jaccard's coefficient $\theta$ (i.e., ${{|E \cap F|}\over{|E \cup F|}} \ge \theta$), where $E$ and $F$ are set-theoretic expressions 
involving $X_i$'s and $Y_i$'s, respectively. For tractability purposes, some syntactic 
bias on the allowable set-theoretic expressions or their length is 
assumed to be provided. For instance, we might restrict $E$ to only involve intersections
of two descriptors from $X$ and $F$ to either an intersection or difference of two
descriptors from $Y$.
Redescription mining hence involves constructive induction (the task of inventing 
new features) and exhibits traits of both unsupervised and supervised learning. It is 
unsupervised because it finds conceptual clusters underlying data, and it can 
be viewed as supervised because clusters defined using descriptors are given meaningful 
characterizations (in terms of other descriptors).

Why is this problem relevant? We posit that today's high-throughput
data-driven sciences are drowning in not just the dimensionality of data, but also 
in the multitude of descriptors available
for characterizing data. Consider gene expression studies using bioinformatics
approaches. The universal set of genes in a given organism ($O$) can be studied
in many ways, such as functional categorizations,
expression level quantification using microarrays, protein interactions, 
and biological pathway involvement.
Each of these methodologies provides a different vocabulary to define subsets of $O$
(e.g., `genes localized in cellular compartment nucleus,' `genes up-expressed two-fold or 
more in heat stress,' `genes encoding for proteins that form the Immunoglobin complex,' 
and `genes involved in glucose biosynthesis'). While traditionally we would
custom-build data mining algorithms to work with each of these vocabularies,
redescription mining provides a uniform way to characterize and analyze the results 
from any of them. In addition, it helps bridge diverse experimental
methodologies by uniformly relating subsets across the corresponding vocabularies.

We further argue that redescription mining serves as a fundamental building block of many
important steps in the iterative, often unarticulated, knowledge discovery process.
A shift of vocabulary allows a given subset of data to be interpretable in a different
context, and allows us to harness existing knowledge from this other context. For
instance, if we
are able to redescribe results from a new stress experiment onto, say, a heat
shock experiment studied earlier, we will be able to study the new results in terms
of known biological knowledge about heat shock.
Chains of redescriptions allow us to relate diverse vocabularies, through important
intermediaries. 

Even redescriptions that hold with Jaccard's coefficient $< 1$ find application in
many domains. An approximate redescription implies a common meeting ground for two
concerted communities of objects. A chain of such approximate redescriptions can 
effectively relate two subsets that have nothing in
common! This is especially useful in {\it story telling} and {\it link analysis} applications.
A query such as `what is the relationship between people traveling on Flight 847 and 
the top 10 wanted list by the FBI?' can be posed in terms of redescription
finding.

While related problems have been studied in the data mining community (most
notably, conceptual clustering~\cite{fisher,michalski}, niche finding, and profiling 
classes~\cite{profiling-classes}), we believe 
that the above formulation of redescription mining has not been attempted before. 
Our contributions here are both the introduction of this new data mining problem, 
as well as a novel tree-based algorithm for mining redescriptions. 

\section{Redescription Mining as Alternating Tree Induction}
We now introduce an approach (CARTwheels) to mining redescriptions that
involves growing two trees in opposite directions, so that 
they are matched at their leaves. The decision conditions in the first tree (say, top)
are based on set membership checks in entries from $X$ and the bottom tree is 
based on membership checks in entries from $Y$; thus
matching of leaves corresponds to a potential redescription. This idea hence
uses paths in the classification trees as representations of boolean expressions
involving the descriptors. 

The CARTwheels algorithm is an alternating
algorithm, in that the top tree is initially
fixed and the bottom tree is grown to match it.
Next, the bottom tree is fixed, and the top tree is re-grown. This process
continues, spouting redescriptions along the way, until designated stopping
criteria are met.

\subsection{Working Example}
For ease of illustration, consider the artificial example in 
Fig.~\ref{eg-data} that shows two sets of descriptors for the universal 
set $O = \{o_1, o_2, o_3, o_4, o_5\}$. 
Here, the set
$X$ corresponds to the set of descriptors $\{X_1, X_2, X_3, X_4\}$ and $Y$ corresponds to
$\{Y_1, Y_2, Y_3, Y_4\}$. The cardinalities of $X$ and $Y$ may not be the same in
the general case. Further, in a realistic application, the number of descriptors would 
far exceed the number of objects.

\begin{figure}
\centering
\begin{tabular}{lll}
\begin{tabular}{l|l|l|l|l|l}
object & $Y_1$ & $Y_2$ & $Y_3$ & $Y_4$ & class \\ \hline
$o_1$  & $\surd$ & $\times$ & $\times$ & $\surd$ & $X_4$ \\
$o_2$  & $\surd$ & $\surd$ & $\times$ & $\surd$ & $X_1$ \\
$o_3$  & $\times$ & $\surd$ & $\surd$ & $\times$ & $X_1$ \\
$o_4$  & $\times$ & $\surd$ & $\times$ & $\times$ & $X_2$ \\
$o_5$  & $\times$ & $\times$ & $\surd$ & $\surd$ & $X_4$ \\ \hline
\end{tabular}
& \hspace{0.5in} &
\begin{tabular}{c} 
\mbox{\psfig{figure=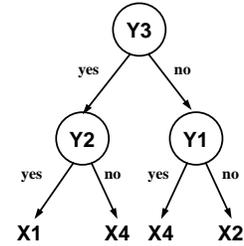,width=1.2in}}
\end{tabular}
\end{tabular}
\caption{(left) Dataset to initialize CARTwheels algorithm. (right)
induced classification tree.}
\label{iter1}
\end{figure}

To initialize the CARTwheels alternation, we prepare a traditional dataset for
classification tree induction, where the entries correspond to the objects,
the boolean features are derived from one of $X$ or $Y$, and the classes
are derived from the other. In the dataset shown in Fig.~\ref{iter1} (left), the
features correspond to set membership in entries of $Y$ 
and each object is assigned a {\it unique} class, chosen from the $X_i$'s it participates
in. We employed a greedy set covering of the objects using the entries of $X$
in order to establish the class labels in Fig.~\ref{iter1} (left).
For instance, $o_2$ belongs to both $X_1$ and $X_3$, but the tie is broken 
in favor of $X_1$. Notice that in this process, $X_3$ does not receive any 
representation in the prepared dataset.  

\begin{figure}
\centering
\begin{tabular}{llll} 
\begin{tabular}{l}
\mbox{\psfig{figure=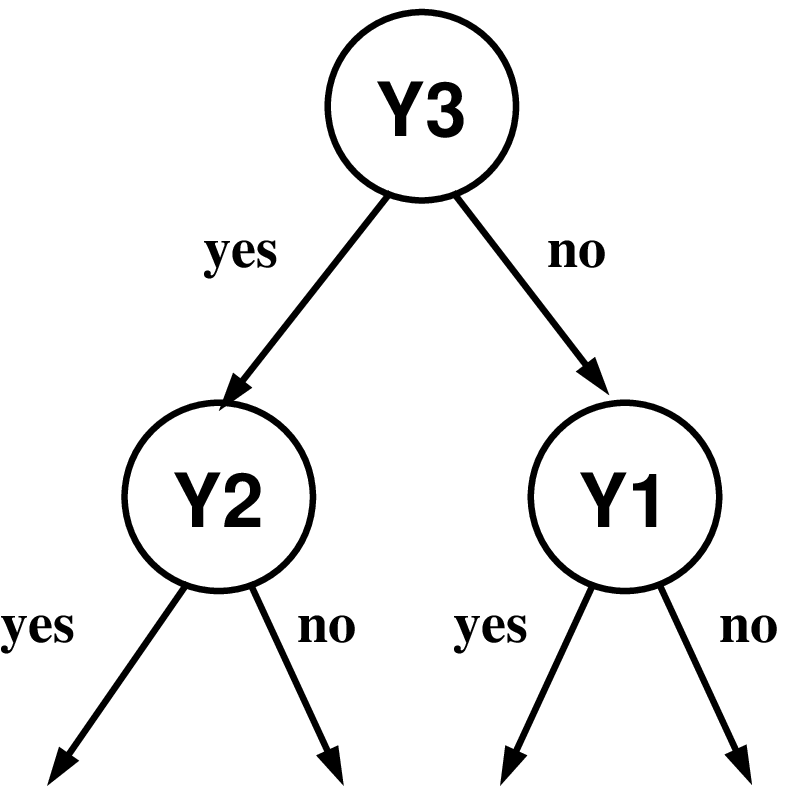,width=1.2in}}\\
\vspace{0.8in}
\end{tabular} &
\begin{tabular}{l}
\mbox{\psfig{figure=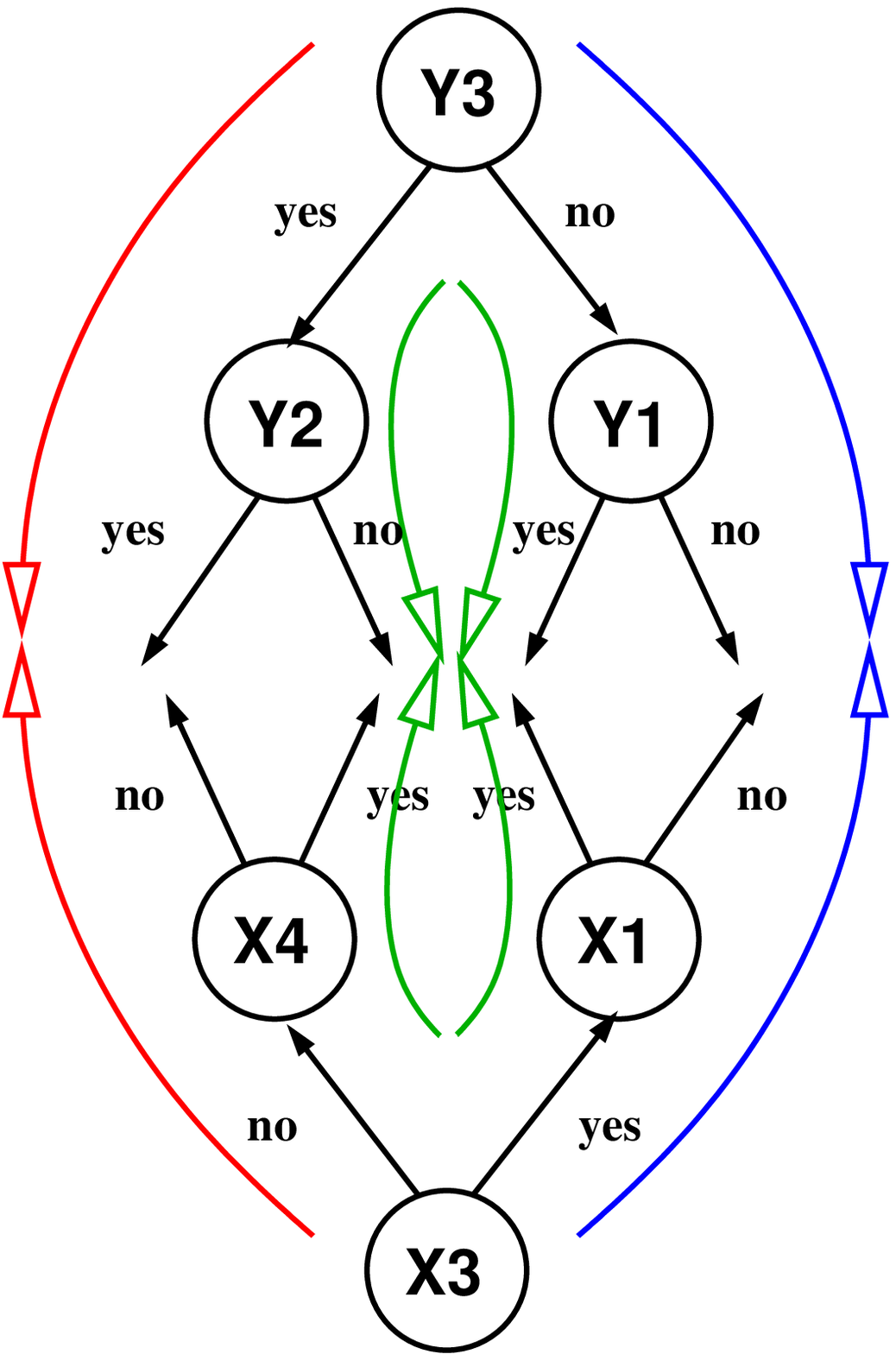,width=1.5in}}
\end{tabular} &
\begin{tabular}{l}
\mbox{\psfig{figure=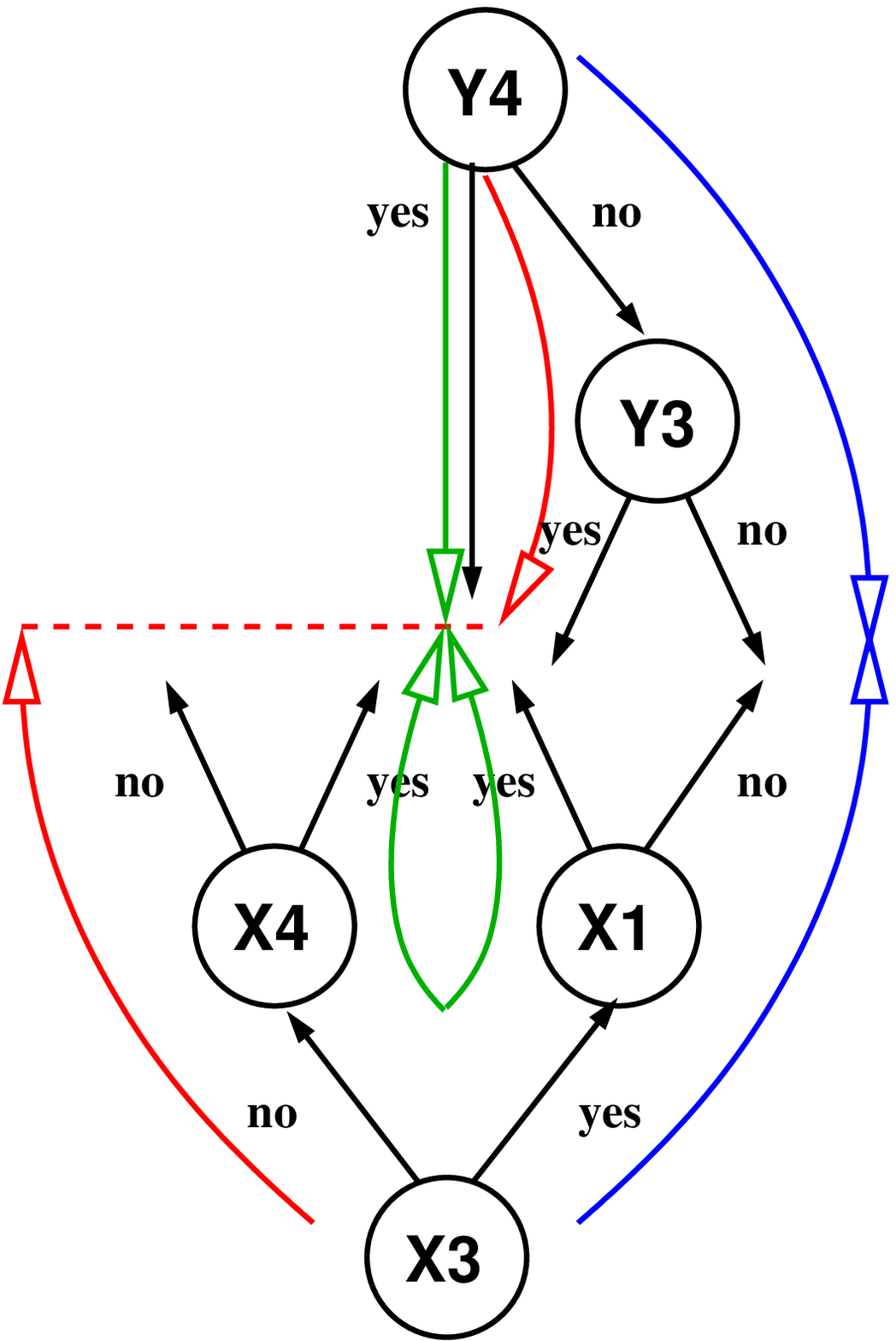,width=1.5in}}
\end{tabular} &
\begin{tabular}{l}
$\cdots \cdots \cdots$
\end{tabular} 
\end{tabular}
\caption{Alternating tree growing in the CARTwheels algorithm. The alternation
begins with a tree (first frame) defining set-theoretic expressions to be matched. 
The bottom tree is then grown to match the top tree (second frame), which is then
fixed, and the top tree is re-grown (third frame). 
Colored arrows indicate the matching paths.
Redescriptions corresponding to matching paths at every stage are read off and subjected to
evaluation by Jaccard's coefficient.}
\label{cart1}
\end{figure}

A classification tree can now be grown using any of the impurity measures studied
in the literature (e.g., entropy, Gini index, misclassification rate). Fig.~\ref{iter1}
(right) depicts a possible tree. The leaves of the tree deterministically predict a class
label from $X$, typically the majority class. At this point, the specific details
of how the tree was induced are not important, only that any such tree will induce
a partition of the underlying objects. In this case, the tree induces a 3-partition
which mirrors the 3-class partition present in the original dataset, but is not
exactly the same. The left most path corresponds to the region $Y_3 \cap Y_2$,
the right most path corresponds to $O - Y_3 - Y_1$, and the union of the two
middle paths gives $(Y_3 - Y_2) \cup (Y_1 - Y_3)$. The reader can verify that these
regions do not have a one-to-one correspondence with the
regions $X_1$, $X_2$, and $X_4$ in the original partition. For instance,
only $X_2$ enjoys such a correspondence, with $O - Y_3 - Y_1$.
In `reading off' a partition from a tree in this manner, 
a conjunction thus results from a path of length $> 1$, a disjunction
results from multiple paths predicting the same class, with negations corresponding
to following the `no' branch from a given node. 
This partition is used as the starting point for the alternation (Fig.~\ref{cart1},
first frame).
\begin{figure}
\centering
\begin{tabular}{lr}
\begin{tabular}{lclllllll}
$X_1$ & = & $\lbrace$ & & $o_2$,& $o_3$ & & & $\rbrace$\\
$X_2$ & = & $\lbrace$ & & & $o_3$, & $o_4$ & & $\rbrace$\\
$X_3$ & = & $\lbrace$ & & $o_2$,& & $o_4$ & & $\rbrace$\\
$X_4$ & = & $\lbrace$ & $o_1$, & & & & $o_5$ & $\rbrace$
\end{tabular}
&
\begin{tabular}{lclllllll}
$Y_1$ & = & $\lbrace$ & $o_1$, & $o_2$,& & & & $\rbrace$\\
$Y_2$ & = & $\lbrace$ & & $o_2$, & $o_3$, & $o_4$ & & $\rbrace$\\
$Y_3$ & = & $\lbrace$ & & & $o_3$, & & $o_5$ & $\rbrace$\\
$Y_4$ & = & $\lbrace$ & $o_1$, & $o_2$, & & & $o_5$ & $\rbrace$
\end{tabular} \\
\end{tabular}
\caption{Example data for illustrating operation of CARTwheels algorithm.}
\label{eg-data}
\end{figure}

\begin{figure}
\centering
\begin{tabular}{ll}
\begin{tabular}{l|l|l|l|l|l}
obj. & $X_1$ & $X_2$ & $X_3$ & $X_4$ & class \\ \hline
$o_1$  & $\times$ & $\times$ & $\times$ & $\surd$ & $(Y_3 - Y_2) \cup (Y_1 - Y_3)$\\
$o_2$  & $\surd$ & $\times$ & $\surd$ & $\times$ & $(Y_3 - Y_2) \cup (Y_1 - Y_3)$\\
$o_3$  & $\surd$ & $\surd$ & $\times$ & $\times$ & $Y_3 \cap Y_2$\\
$o_4$  & $\times$ & $\surd$ & $\surd$ & $\times$ & $O - Y_3 - Y_1$\\
$o_5$  & $\times$ & $\times$ & $\times$ & $\surd$ & $(Y_3 - Y_2) \cup (Y_1 - Y_3)$\\ \hline
\end{tabular} &
\begin{tabular}{l|l|l|l|l|l}
obj. & $Y_1$ & $Y_2$ & $Y_3$ & $Y_4$ & class \\ \hline
$o_1$  & $\surd$ & $\times$ & $\times$ & $\surd$ & $(X_3 \cap X_1) \cup (X_4 - X_3)$ \\
$o_2$  & $\surd$ & $\surd$ & $\times$ & $\surd$ & $(X_3 \cap X_1) \cup (X_4 - X_3)$ \\
$o_3$  & $\times$ & $\surd$ & $\surd$ & $\times$ &  $(O - X_3 - X_4)$ \\
$o_4$  & $\times$ & $\surd$ & $\times$ & $\times$ & $(X_3 - X_1)$\\
$o_5$  & $\times$ & $\times$ & $\surd$ & $\surd$ &  $(X_3 \cap X_1) \cup (X_4 - X_3)$\\ \hline
\end{tabular}
\end{tabular}
\caption{(left) Dataset for second iteration of CARTwheels algorithm. Notice that
class labels are now set-theoretic expressions involving $Y_i$'s. (right)
Dataset for third iteration of CARTwheels algorithm.}
\label{iter2}
\end{figure}

We now prepare a dataset with entries from $X$ as the features
and the regions thus formed (involving $Y_i$'s) as the classes, as
shown in Fig.~\ref{iter2} (left). Inducing a classification tree from this dataset
really corresponds to growing a second tree to match the first tree at the
leaves, as depicted in Fig.~\ref{cart1} (second frame). In this case, the second tree
also learns a 3-partition and we can evaluate each of these matchings using 
the Jaccard's measure. This produces three redescriptions: 

\vspace{-0.2in}
\begin{eqnarray*}
(X_3 \cap X_1) \cup (X_4 - X_3) & \Leftrightarrow & (Y_3 - Y_2) \cup (Y_1 - Y_3) \\
(X_3 - X_1) & \Leftrightarrow & (O - Y_3 - Y_1) \\
(O - X_3 - X_4) & \Leftrightarrow & (Y_3 \cap Y_2) \\
\end{eqnarray*}

\vspace{-0.3in}
\noindent
all of which hold at Jaccard's coefficient $1$. This need not 
be the case in general. The bottom tree might be able
to match only some paths in the top tree, or the matches might not pass our
Jaccard's cutoff. This process is then continued, now with $Y_i$'s as features
and the partitions derived from the bottom tree as classes (see
right of Fig.~\ref{iter2}). The new matchings yield the redescriptions:

\vspace{-0.2in}
\begin{eqnarray*}
(X_3 \cap X_1) \cup (X_4 - X_3) & \Leftrightarrow & Y_4 \\
(O - X_3 - X_4) & \Leftrightarrow & (Y_3 - Y_4) \\
(X_3 - X_1) & \Leftrightarrow & (O - Y_3 - Y_4)\\
\end{eqnarray*}

\vspace{-0.3in}
\noindent
which, fortuitously, also have a Jaccard's coefficient of $1$. Notice that, this time, the
root decision node that has been picked is $Y_4$ (see third frame
of Fig.~\ref{cart1}) and the tree actually
resembles a {\it decision list} (a tree where every internal node has a leaf on 
its `yes' branch). The alternation can be continued (see next section for ways to configure the search).

If we limit the size of the trees at every iteration, it is easy to see that the
set-expressions constructed cannot get arbitrarily long.
In our running example, we use a depth limit of 2 so that all expressions on 
either side of a mined redescription can involve at most three descriptors. The
longest expressions result from unions of two paths involving different subtrees. 

\subsection{The CARTwheels Algorithmic Framework}
Why does CARTwheels work? The use of trees to mine one-directional
implications (rules) is well understood and is the idea behind algorithms such
as C4.5~\cite{c45}. In CARTwheels, we exploit the duality between {\it class partitions}
and {\it path partitions} to posit the stronger notion of equivalence. In fact,
if a tree reduces the entropy to zero, it is clear that there must be a one-to-one
correspondence between its path partitions and class partitions, which are really
path partitions from the other tree. Keep in mind that different paths are union-ed 
when they predict the same class, and this property is crucial to establishing the duality.

The search for redescriptions in CARTwheels can be viewed as a problem of identifying
(and creating)
correlated random variables. A descriptor, e.g., $D$, can be considered to be
a discrete random variable that takes on values from $O$. Every object in $D$ occurs
with probability ${{1}\over{|D|}}$ and other objects occur with probability zero, to
yield total probability mass of $1$. Notice that this makes the self entropy
of such a random variable to be the logarithm of the size of the descriptor.
Now consider
running a CARTwheels alternation with a depth limit of 1 for the classification
trees. Mining a redescription with Jaccard's coefficient of $1$ means that 
we have identified a random variable $D'$ whose {\it entropy distance} from
$D$ is zero. The entropy distance~\cite{mackay} is given by:
$$H(D, D') - I(D; D')$$
where $H(D, D')$ is the joint entropy function of \{$D$, $D'$\}
and $I$ qualifies the mutual information, in turn
given by:
$$I(D; D') = H(D) - H(D|D')$$
\noindent
where $H(D)$ is the self-entropy of $D$ and $H(D|D')$ is the conditional entropy
of $D$ given $D'$.
In other words, the average reduction in uncertainty about $D$ due to knowing
$D'$ is exactly the self entropy of $D$, causing an entropy distance of $0$.
Entropy distance is a true distance measure, unlike measures such as the
Kullback-Leibler (KL) divergence. Smaller values of entropy distance hence imply
higher values of Jaccard's coefficient.

When we increase the depth limit, the analysis gets complicated
because the redescription mining problem as stated is severely underconstrained. 
Any successful algorithm is required to arrive at both the set theoretic expressions 
as well as test them for equivalence. The growing of classification trees with boolean 
features multiplexes the steps of constructive induction and guarantees implication, while
the duality of partitions helps set the stage for equivalence testing. 
Ideally we would like to exercise precise control over the sequence in which the 
algorithm explores options, in order to qualify the nature of the mined redescriptions.

\vspace{-0.1in}
\subsubsection{Modeling CARTwheels Alternation}
Towards this end, the operation of CARTwheels can be modeled as a Markov process since the choice of next
state is a function of only the current state (and perhaps global information such as
$O$, $X$, and $Y$). This means that we can reduce the search for potential
redescriptions to the design of a suitable state exploration policy.

What does `state' mean in modeling the operation of CARTwheels? Refer back to
the duality of path partitions and class partitions -- either of them could be used
as a representation of state. The representation can be given as
a label vector employing some canonical ordering of the objects in $O$
(to ensure uniqueness). In addition, we have the option of including the
descriptors for these partitions (in terms of $X_i$'s or $Y_i$'s) as part
of the state representation. This is important when the same partition 
is realized by different descriptors, but must be considered distinct
for redescription mining purposes (a simple example arises when 
$X$ or $Y$ have elements that are exactly the same). We employ
this approach in our studies.

\vspace{-0.1in}
\subsubsection{Designing an Exploration Policy}
Once the representational issue is decided, the more fundamental question
pertains to the design of a suitable exploration policy. In contrast to
traditional classification tree induction which is motivated at
{\it reducing} entropy, CARTwheels must actually {\it maintain} entropy in
some form, since impurity drives exploration. 

Should CARTwheels attempt to find {\it all} redescriptions? This is clearly
a tall order, and is reminiscent of the difficulties encountered in 
association rule mining~\cite{srikant-association}, where the number of rules generated can
quickly become unwieldy. Attempting to do this in the CARTwheels framework
is unappealing since every set expression postulated by one tree must 
be matched by every expression modelable in the other tree! This will require
multiple visitations of the same state and, while we can interleave the
testings for matches to a certain extent, would involve enormous overhead in
book-keeping.  Instead we can exploit the algebraic structure of
the problem to identify a minimal generating set of redescriptions, and design
the policy to only visit the relevant states for this purpose. This is similar
to the strategy pursued by Zaki for mining a non-redundant set of association
rules~\cite{zaki-non-redundant}.
How this can be done effectively for redescriptions is the topic of a future paper. 

A second approach forsakes the desire to explore all redescriptions, and
instead exploits the property that a redescription can be viewed as a subset
of $O \times O$ space, i.e., a binary relation on $O$. Here, instead of computing 
all possible equivalences, we only find enough redescriptions to cover this space 
a specified number of times. We have to be careful here because redescriptions occur 
in two flavors. A redescription with Jaccard's coefficient $1$ is a strong one, and
has a {\it complementary redescription}---with both left side and right side expressions
negated---that will also be strong. A redescription with Jaccard's coefficient
$< 1$ is approximate and might hold in only one complement.
For instance, if both sides of the redescription
cover, say, 90\% of the objects in $O$, then a very high Jaccard's coefficient
can result purely by chance! Needless to say, the complementary redescription involving
the remaining 10\% of the objects may not hold.
The net effect is that some redescriptions might imply a complete cover of
$O \times O$ space, whereas others will only cover subsets. A workable
criterion of coverage hence requires careful study.

In this paper, we employ a simpler exploration policy where descriptors
participating in a path (but not the leaf) and yielding a good redescription
are removed from consideration in subsequent alternations. In experimental
tests, we have observed that this greedy policy guarantees a rapid exhaustion 
of the sets $X$ and $Y$. We also place a limit on the
number of alternations that the algorithm can pass through without
finding any redescriptions. Once this limit is exceeded (happens after many useful
entries from $X$ and $Y$ have been deleted), the algorithm terminates.

CARTwheels also employs randomization heuristics to facilitate state exploration.
For instance, when assigning class labels after inducing a tree, we take care to ensure
that the same label is not assigned to all leaves, and suitably randomize assignments
toward this purpose. This might appear counter-intuitive, but notice that
it only has the effect of re-organizing the partitions derived from the paths, and
can be seen as buying time for the Markov chain. Another example pertains to how 
decision nodes are selected for inclusion in a tree. While we use entropy as the 
primary driver for tree induction, we sometimes perform randomized moves at the 
root level, in order to prevent over-dominance of one descriptor in the ensuing 
redescriptions. 

\subsection{Implementation Details}
CARTwheels is implemented in C++ atop a Postgres database providing access
to the descriptors. We use an AD-tree data structure~\cite{moore-lee} for 
fast counting purposes and estimation of entropy (this is distinct from the
classification tree that combines the descriptors). 
The AD-tree provides access to the distributions of `class labels' for every combination 
of `features' and, since the definition of features and class labels change at
every iteration, is rebuilt continually. Notice that the data
structure is expected to provide
both the sizes of descriptors as well as their negations (when we follow
the `no' branch) and hence, the depth of the AD-tree is set to just greater than the 
allowable depth of the classification trees. The CARTwheels algorithm consults
the AD-tree whenever it must make a choice of a decision node (except when its move
is exploratory). After evaluating matchings, set-expressions read off the trees
are subjected to tabular minimization, in order to arrive at a canonical form.

The implementation allows for configuring the space of redescriptions that are explored.
The depth limit for the top and bottom trees can be individually specified, and 
we can also preferentially include or exclude certain types of expressions in
mined redescriptions. For instance, syntactic constraints on redescriptions (e.g., only
conjunctions are allowed) can be incorporated as biases in the tree construction 
phase of CARTwheels.

\section{Applications in Bioinformatics}
\label{app}
We now present an application of CARTwheels to studying gene expression datasets 
from microarray experiments conducted on the budding yeast {\it Saccharomyces cerevisiae}.
Bioinformatics is fertile ground for application of CARTwheels and
{\it S. cerevisiae} is arguably the most well studied (and documented)
model organism through bioinformatics techniques.
Practically every experimental methodology applied towards yeast 
can be viewed as a way to define descriptors. 
Even the results of other data analysis/mining algorithms can be used as a source 
of descriptors! The underlying universal set of objects could be initialized to 
the set of genes, proteins, or processes, in {\it S. cerevisiae}.
CARTwheels hence brings many computational and experimental technologies 
to bear upon redescription mining. It supports the capture of
both similarities and distinctions among descriptors derived from these diverse
sources.

The redescription process begins by defining an universal set of genes
$O$, which is dependent on our biological goals. Here, we are interested in 
characterizing similarities and differences in yeast gene expression behavior across 
related families of stresses. 
Gasch et al.~\cite{gasch} is an important source
for such a study since it provides results from more than 170 comparisons, across
a variety of environmental stresses. We selected five stresses from this dataset
(heat shock from $25^\circ$C to $37^\circ$C,
hyper-osmotic shock, hypo-osmotic shock, $H_2 O_2$ exposure, and mild heat shock at 
variable osmolarity) and initialized $O$ to be the set of genes that show
significant expression (more than 1-fold up- or down-expression)
in some time point in each of these stresses. This results in a set of 74 genes/ORFs.

The choice of the universal set can be viewed as a conditioning context and must
be kept in mind when interpreting any mined redescriptions. It can be viewed as
an implicit descriptor occurring on both sides of every redescription, e.g.,
$E \Leftrightarrow F$ in $O$ can be viewed as $E \cap O \Leftrightarrow F \cap O$.

We defined 824 descriptors, in a variety of ways. One class of descriptors
was derived from categories in the GO biological process, GO cellular component, and 
GO molecular function taxonomies, that have representation among the chosen 74 genes. 
This yields a total of 378 descriptors (210 GO BIO + 42 GO CELL + 126 GO MOL). 
The microarray results from the five stresses of Gasch et al.~\cite{gasch} were
bucketed to yield range descriptors of 
the form `expression level $\in $ [\%x, 0]
in time point \%y of stress experiment \%z' (for negative \%x) and
`expression level $\in $ [0, \%x] in time point \%y of stress experiment \%z' (for 
positive \%x). This produces 224 descriptors.
Further, k-means clustering
was performed using the Genesis software suite~\cite{genesis} on each of the 
stresses individually, with a setting of 10 clusters. Since heat
shock and mild heat shock at variable osmolarity are actually pairs of
experiments, this step yields (5+2) $\times$ 10 = 70
descriptors depicting clusters of genes with 
similar time profiles. Finally, we included microarray
results from a histone depletion experiment conducted by Wyrick et al.~\cite{wyrick}
and created 152 range descriptors similar to the Gasch stresses; this is to allow
us to relate the effect of histone depletion to that of
environmental stresses.

To invoke CARTwheels, we initialized $X$ to be all descriptors derived from 
the Gasch et al. dataset (which includes the range descriptors as well as the k-means
clusters). This ensures that all redescriptions will involve some aspect of the Gasch
et al. experiment and prevents the possibility of, say,
mining a redescription between two GO taxonomies.
$Y$ was initialized to the set of all descriptors; thus, there is some overlap between
$X$ and $Y$. In order to prevent obvious redescriptions arising from this overlap,
the algorithm was precluded from utilizing descriptors in one tree if they are already
present in the other tree. 

We employed a Jaccard's threshold of 0.5 and a depth-limit of 2 in both the top
and bottom tree induction alternations. The limit on the number of allowable
alternations till a redescription is mined is set to 10.
Redescriptions inferred from CARTwheels
are required to hold in both the mined and complementary forms. For example,
for the equivalence $E_1 \cup E_2 \Leftrightarrow F$ to be considered as a
redescription, it must hold with Jaccard's coefficient at least 0.5, as 
must its complement: $\neg E_1 \cap \neg E_2 \Leftrightarrow \neg F$.
This ensures that every redescription truly induces a partition of $O$.
Thus whetted,
redescriptions are then subjected to a `tightening' step, akin to rule pruning
in packages like C4.5. This might involve attempting to drop terms from 
both sides of the redescription, or restricting range descriptors (if they
occur in the redescription),
and determining whether this causes significant degradation of Jaccard's
coefficient. If no degradation is observed, then the redescription can be tightened.
With these design choices, and the greedy exploration policy,
CARTwheels terminates after using 150 of the 824 descriptors, yielding about 
200 redescriptions. 

\begin{figure}
\centering
\begin{tabular}{l}
\includegraphics[width=6.3in]{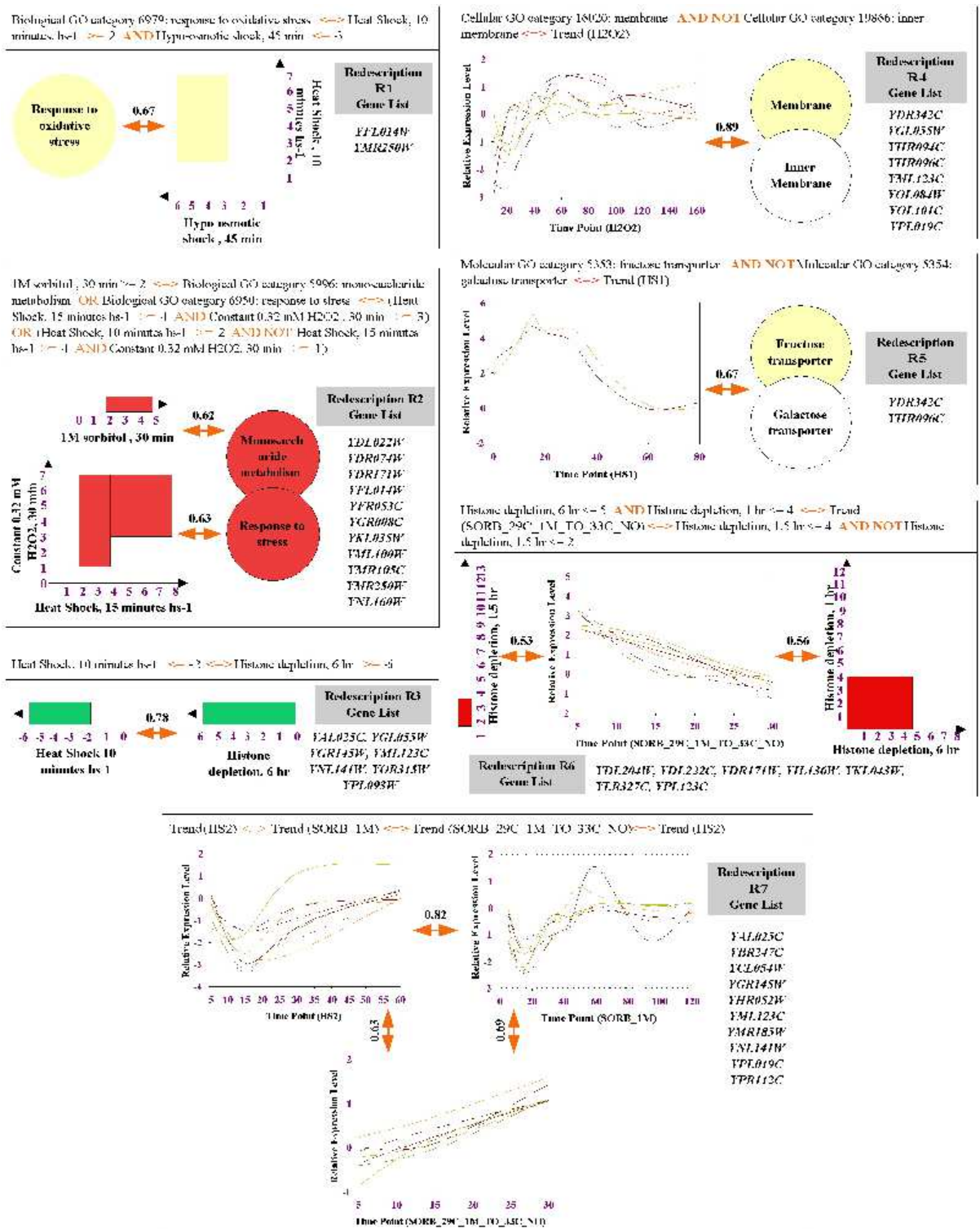}
\end{tabular}
\caption{Seven redescriptions mined from gene expression studies on
{\it Saccharomyces cerevisiae}. Each box gives a readable statement of the
redescription, presents it in graphical form, and identifies the genes
conforming to the redescription. The Jaccard's coefficient is displayed over
the redescription arrow. Notice that some redescriptions (e.g., R1, R5) involve
only two genes, whereas others such as R2 involve larger numbers.}
\label{redescs-final}
\end{figure}

Seven key mined redescriptions (R1--R7) are depicted in Fig.~\ref{redescs-final}. 
They were selected for both their biological interest as well as for their 
feature construction novelties. R1 is a redescription between the GO taxonomy
and experimental stresses from the Gasch dataset, and involves two genes. The 
rectangular region
on the right side is bounded by the extremal values specific to the experiment,
and hence is captured by a conjunction of merely two descriptors. From a biological
perspective, R1 is interesting because it relates {\it contrasting} behavior in
two different experimental comparisons (positive in heat shock 10 minutes,
and negative in hypo-osmotic shock) to a GO biological category related to stress.

R2 is actually a chain of two redescriptions, mined in successive iterations of
CARTwheels. This redescription involves 11 genes and relates the disjunction of
two different GO biological categories to expression data across three different
stresses (this time, all involving positive expression). Notice that one of the
derived expression descriptors is also a disjunction. It is pertinent to note
that R2 experienced some tightening of its range descriptors -- this is why
one of its expressions has an extra term than would be expected for a 2-level
tree.

R3 satisfies our curiosity about the similarity between the histone depletion
experiment and a Gasch comparison (heat shock). It involves 7 genes, two of
which are hypothetical and one with a putative annotation. Such redescriptions
involving un-annotated genes are important for suggesting testable hypotheses
about their functionality.

\newpage
The remaining redescriptions involve cluster profiles on one or both sides.
R4 relates a k-means cluster to a set difference of two {\it related} GO cellular
component categories. Interestingly, two of the genes participating in this redescription
(YDR342C and YHR096C) are singled out by the next redescription (R5), which 
identifies a different k-means cluster to characterize these genes; and which also uses
a set-difference, this time of related GO molecular function categories. 

R6 is another chain of redescriptions, similar to R2, and relates a particular
trend in the Gasch dataset to positively expressed genes in three different time
points of the histone experiment. It involves 7 genes. Finally, R7 is 
actually a triangle of redescription relationships that illustrates the power of
CARTwheels. Three different experimental comparisons are involved in this
circular chain of redescriptions, with 10 genes being implicated in all three
descriptors. From a biological standpoint, this is a very interesting result --
the common genes indicate concerted participation across stress conditions; whereas
the genes participating in, say, two of the descriptors, but not the third, suggest 
a careful diversification of functionality.

\section{Discussion}
This paper is a first exploration into the formulation of the redescription
mining problem and has presented an approach for mining redescriptions automatically.
Redescriptions can be thought of as generalizations of one-directional 
implications (e.g., association rules~\cite{srikant-association}, rules in ILP~\cite{muggleton-ilp}), where
one descriptor is required to be a proper subset of the other. This 
generalization coupled with the automatic
identification of set-theoretic constructions makes CARTwheels a very powerful
approach to mining (approximate) equivalence relations. 
We have demonstrated the effectiveness of CARTwheels in a domain that 
exhibits a richness of descriptors, and shown how it captures patterns involving
small as well as large sets of objects.

The work presented here can be considered a significant extension of 
ideas pursued in the schema matching~\cite{schema-matching}, clustering categorical 
data~\cite{clustering-categorical}, and model management~\cite{model-management}
literature. The relationships considered in schema matching research are
primarily of the foreign key nature or otherwise operate at the instance level,
whereas we consider more complex set-theoretic
relationships. Clustering categorical data focuses on defining similarity measures
in non-metric spaces and this research can be fruitfully integrated with our work.
However, notice that we are not merely clustering data but also imposing
describability constraints. 
Model management is a framework that recognizes the complex inter-relationships
that would exist in multi-database enterprises and provides union, intersection,
and difference operators for reconciliation, integration, and migration purposes. 
The relationships here are assumed to be user provided, and the emphasis is
on actually `executing a redescription.' CARTwheels can thus be usefully employed here
as a driver for determining what these relationships should be.

We now outline some directions for future research.
The connection between Jaccard's coefficient and algorithmic driver parameters
(such as entropy) deserves further study. Other ways of evaluating 
redescriptions~\cite{dissimilarity,jones-furnas} are also pertinent here
(e.g., Dice coefficient) and some of these could support more efficient 
tree-based algorithms than the Jaccard's coefficient. Ideally, an evaluation metric
would obey some closure properties in the space of redescriptions, which can be used
to configure an exploration strategy. In addition, it is preferable that
an evaluation metric lends itself to the design of a statistical test of
significance for redescriptions.

Thus far, we have assumed a `flat' organization of the given descriptors and do 
not recognize any structural relationships between them. However, some descriptor vocabularies
(e.g., derived from GO) enjoy a hierarchical structure, which can be exploited by
the mining algorithm. Specialized redescription algorithms can thus be designed for
targeted descriptor families.

There is an intrinsic limit to a dataset's potential to reveal redescriptions,
which can be studied through statistical analysis of set size distributions and
estimates of overlap potential. Of particular interest here is
qualifying the `expected' results from a
CARTwheels alternation before actually performing the alternation; 
the {\it entropy rate} of the stochastic process underlying the Markov 
chain~\cite{cover-thomas} can be a useful indicator in this regard.

Our current focus is on using redescriptions to automatically span multiple levels
of abstraction (e.g., gene subsets $\rightarrow$ pathways $\rightarrow$ biological
processes). This would firmly establish the importance of redescription in 
bridging the diverse levels at which information is created and characterized.

\subsection*{Acknowledgements}
This work is supported through the Multidisciplinary University Research Initiative
(MURI) program of the Department of Defense (Biomimetic Cell and Tissue Stasis;
N00014-01-1-0852), the Metabolic Engineering for Cellular Stasis program of
DARPA (N00173-98-1-G005-P00004 and N00173-02-1-G016), and the National Science Foundation
(Next Generation Software program; grants EIA-9984317 and EIA-0103660). We thank
Bud Mishra, Chris Bailey-Kellogg, and T.M. Murali
for critically reading the manuscript and providing useful comments.

\bibliographystyle{plain}
\bibliography{paper}
\end{document}